\documentclass[10pt,letterpaper]{article}
\usepackage{opex3}
\usepackage{subfigure}
\usepackage{graphicx}
\usepackage{epsfig}
\usepackage{cite}
\usepackage{amsmath,bbm}
\newcommand{\ket}[2][]{{|#2\rangle_{#1}}}
\newcommand{\bra}[2][]{{}_{#1}\langle #2|}

\newcommand{\proj}[2][]{\ket{#2}_{#1}\bra{#2}}

\begin{document}

%%%%%%%%%%%%%%%%%% title page information %%%%%%%%%%%%%%%%%%
\title{Scheme for on-chip verification of\\ transverse mode entanglement using the\\  electro-optic effect}
\author{Divya Bharadwaj,$^{1}$ K. Thyagarajan,$^{1}$ Micha{\l} Jachura,$^{2*}$ Micha{\l}~Karpi{\'n}ski,$^{2}$ and Konrad~Banaszek$^{2}$ }

\address{$^{1}$Department of Physics, IIT Delhi, New Delhi 110016, India\\
$^{2}$Faculty of Physics, University of Warsaw, Pasteura 5, 02-093 Warsaw, Poland\\
}

\email{$^{*}$michal.jachura@fuw.edu.pl}



\begin{abstract}
A key ingredient in emerging quantum-enhanced technologies is the ability to coherently manipulate and detect superpositions of basis states. In integrated optics implementations, transverse spatial modes supported by multimode structures offer an attractive carrier of quantum superpositions. Here we propose an integrated dynamic mode converter based on the electro-optic effect in nonlinear channel waveguides for deterministic transformations between mutually non-orthogonal bases of spatial modes. We theoretically show its capability to demonstrate a violation of a Bell-type Clauser-Horne-Shimony-Holt inequality by measuring spatially mode-entangled photon pairs generated by an integrated photon pair source. The proposed configuration, numerically studied for the potassium titanyl phosphate (KTP) material, can be easily implemented using standard integrated optical fabrication technology.
\end{abstract}

 % REPLACE WITH CORRECT OCIS CODES FOR YOUR ARTICLE

%%%%%%%%%%%%%%%%%%%%%%% References %%%%%%%%%%%%%%%%%%%%%%%%%

\section{Introduction}
Realization of spontaneous parametric down-conversion in $\chi^{(2)}$ nonlinear optical waveguides \cite{1,7,2} offers novel strategies to engineer photonic entanglement, which powers numerous applications in the emerging field of quantum-enhanced technologies
\cite{PhotonicQuantumTechnologies}. Entanglement between photons can be produced in various degrees of freedom such as polarization \cite{3,4,5,6}, time/frequency \cite{8,9,10,11,12} etc. In the case of multimode waveguides, transverse spatial modes can be also used as the basis for generating a discrete (finite-dimensional) entangled state \cite{9,MosleyChristPRL09,17,KarpinskiOL12,KarpinskiSPIE, Olomouc}. 

In this paper, we present an electrically-controlled integrated optics scheme to verify photonic entanglement in transverse spatial modes supported by a multimode waveguide. As a radiation source, we employ a high-fidelity maximally entangled state in transverse modes generated using a quasi phase matched (QPM) nonlinear channel waveguide fabricated in potassium titanyl phosphate (KTP). At the heart of our scheme is an original design for dynamic mode converters using the electro-optic effect in two-moded channel waveguides, which enable measurements of generated photons in non-orthogonal spatial bases. These components are employed to analyze the violation of the Clauser, Horne, Shimony, and Holt (CHSH) inequality \cite{13} for spatially entagled photons in analogy to standard tests of polarization entanglement \cite{14,15,16}. The presented results contribute to a universal toolbox for utilizing prospectively the spatial degree of freedom of guided optical radiation in a variety of quantum protocols. Recently, the acousto-optic effect has been used to couple two transverse modes of an optical fiber to verify mode entanglement \cite{AOM}. Our approach offers the possibility to perform deterministic, polarization-selective operations on pairs of spatial modes without radio frequency electric signals.

This paper is organized as follows. Sec.~\ref{Sec:Principle} describes the principle of the scheme. Its quantitative analysis is carried out in Sec.~\ref{Sec:Analysis} and numerical results are presented in Sec.~\ref{Sec:Numerical}. Finally, Sec.~\ref{Sec:Conclusions} concludes the paper.

\section{Principle}
\label{Sec:Principle}
\begin{figure}[b]
\centering\includegraphics[width=13cm]{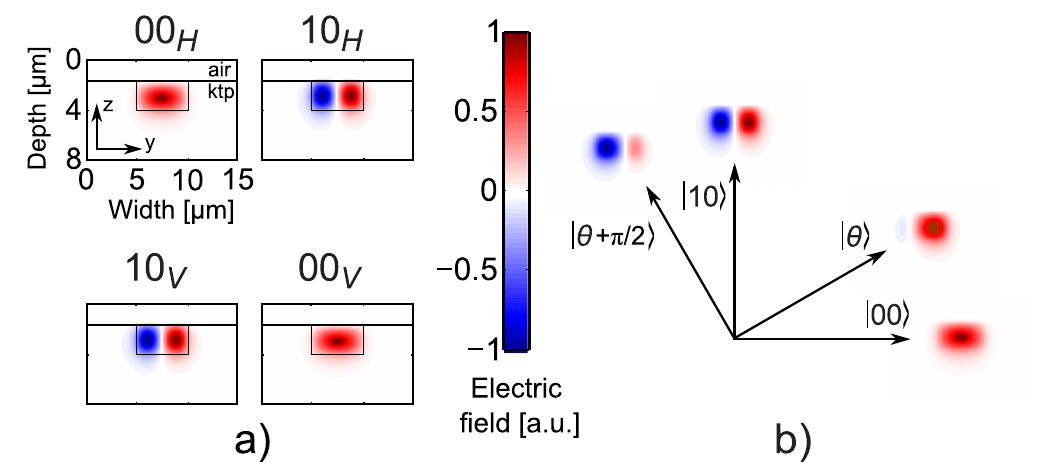}
\caption{(a) Electric field distributions for transverse spatial modes $00$ and $10$ for horizontal $H$ and vertical $V$ polarizations in an exemplary KTP waveguide used for the numerical example presented in Sec.~\ref{Sec:Numerical} of this paper. (b) Rotation of the spatial mode basis required to test the CHSH inequality. Rotated modes $\ket{\theta}$ and $\ket{\theta+\frac{\pi}{2}}$ are illustrated with distributions depicted for $\theta=30^\circ$.}
\label{Fig:1}
\end{figure}

We consider spontaneous parametric down conversion (SPDC) in a multimode channel waveguide fabricated in a $z$-cut, $x$-propagating KTP
substrate; here $xyz$ coordinate system refers to the principal axis system of the crystal. In a type-II process, the two generated photons can be distinguished by their orthogonal polarizations, associated with the signal and the idler beam. We will refer to these polarizations as horizontal $H$ and vertical $V$, with dominant components of the electric field oriented respectively along $y$ and $z$ axes. For each polarization we will be interested in two transverse spatial modes depicted in Fig.~\ref{Fig:1}(a): the fundamental one $00$ and a first-order one $10$ with a node defined by the equation $y=0$.

When a horizontally polarized pump $P$ is prepared in the transverse mode $10_P$, two processes can be simultaneously realized \cite{17}:
\begin{align}
1: & \qquad 10_P \rightarrow 00_H + 10_V \nonumber \\
2: & \qquad 10_P \rightarrow 10_H + 00_V.
\label{Eq:TwoProcesses}
\end{align}
With the right choice of the waveguide geometry, the conversion efficiencies and spectral characteristics of these two processes can be made nearly identical, which produces to a good approximation a pure two-photon state in the form
\begin{equation}
\ket{\Psi}=\frac{1}{\sqrt{2}}\big(\ket{00_{H},10_{V}} + \ket{10_{H},00_{V}}  \big).
\label{Eq:MaxEntState}
\end{equation}
This state can be viewed as a maximally entangled state of two qubits labeled by the polarization indices $H, V$. Each qubit space is spanned by a pair of orthogonal basis states defined by transverse modes $00$ and $10$. The qubits can be separated in a straightforward manner using a polarizing beam splitter and sent in different directions to satisfy the locality condition when testing the CHSH inequality. However, to demonstrate the flexibility of manipulating such spatial qubits in wave guiding structures we will consider independently controlled transformations of $H$ and $V$ photons travelling along the same path.

The state introduced in Eq.~(\ref{Eq:MaxEntState}) can serve as a resource to test the violation of the CHSH inequality analogously to standard polarization entanglement. In the case of a polarization entangled state, verification of entanglement is carried out by passing the two photons through two independent polarizers (or polarization splitters) and by changing the orientation $\theta$ of their pass axes followed by measuring coincidences between the two detectors \cite{16}. Basically, the rotation of the axes of the polarizers (or polarization splitters) results in a mixing of the two orthogonal polarization components or equivalently in projecting the photons onto non-orthogonal basis states. By choosing an appropriate combination of the projections on the two photons it is possible to show the violation of CHSH inequality and thus verify polarization entanglement.

In order to extend this idea to demonstrate spatial mode entanglement for photon pairs described by Eq.~(\ref{Eq:MaxEntState}), we need to implement projections onto general superpositions of the form
\begin{align}
\ket{\theta} & = \cos\theta \ket{00} + \sin\theta \ket{10} \nonumber \\
\ket{\theta+{\textstyle\frac{\pi}{2}}} & = - \sin\theta \ket{00} + \cos\theta\ket{10},
\end{align}
shown schematically in Fig.~\ref{Fig:1}(b).
This can be achieved using single-qubit gates that would realize a unitary transformation
$\ket{\theta} \rightarrow \ket{00}$ and $\ket{\theta+{\textstyle\frac{\pi}{2}}} \rightarrow \ket{10}$, followed by detection in the basis $\{\ket{00}, \ket{10}\}$. A device implementing such a gate is a dynamic mode converter which controllably couples the two spatial modes with the angle $\theta$ defined by the coupling strength between the two modes.

We define the difference between the propagation constants of the two spatial modes in $H$ and $V$ polarizations that are to be coupled by $\Delta\beta_{H(V)} = \beta^{H(V)}_{00}-\beta^{H(V)}_{10}$, where $\beta^{H(V)}_{mn}$ are the propagation constants of $H(V)$-polarized $mn$ modes of two-mode channel waveguide ($m$ = 0, 1; $n$ = 0). Since $\Delta\beta_{H(V)}$ is non-zero, we need two consecutive periodic electrode patterns with appropriate spatial frequencies $K_{H(V)} = \frac{2\pi}{\Lambda_{H(V)}}$ to couple the two different $H(V)$-polarized modes 00 and 10; for maximum efficiency we must have $K_{H(V)} = \Delta\beta_{H(V)}$ \cite{18}. Here $\Lambda_{H(V)}$ are the spatial periods of the periodic electrode patterns designed to couple $H$- or $V$-polarized modes respectively.

\begin{figure}[b]
\centering\includegraphics[width=8cm]{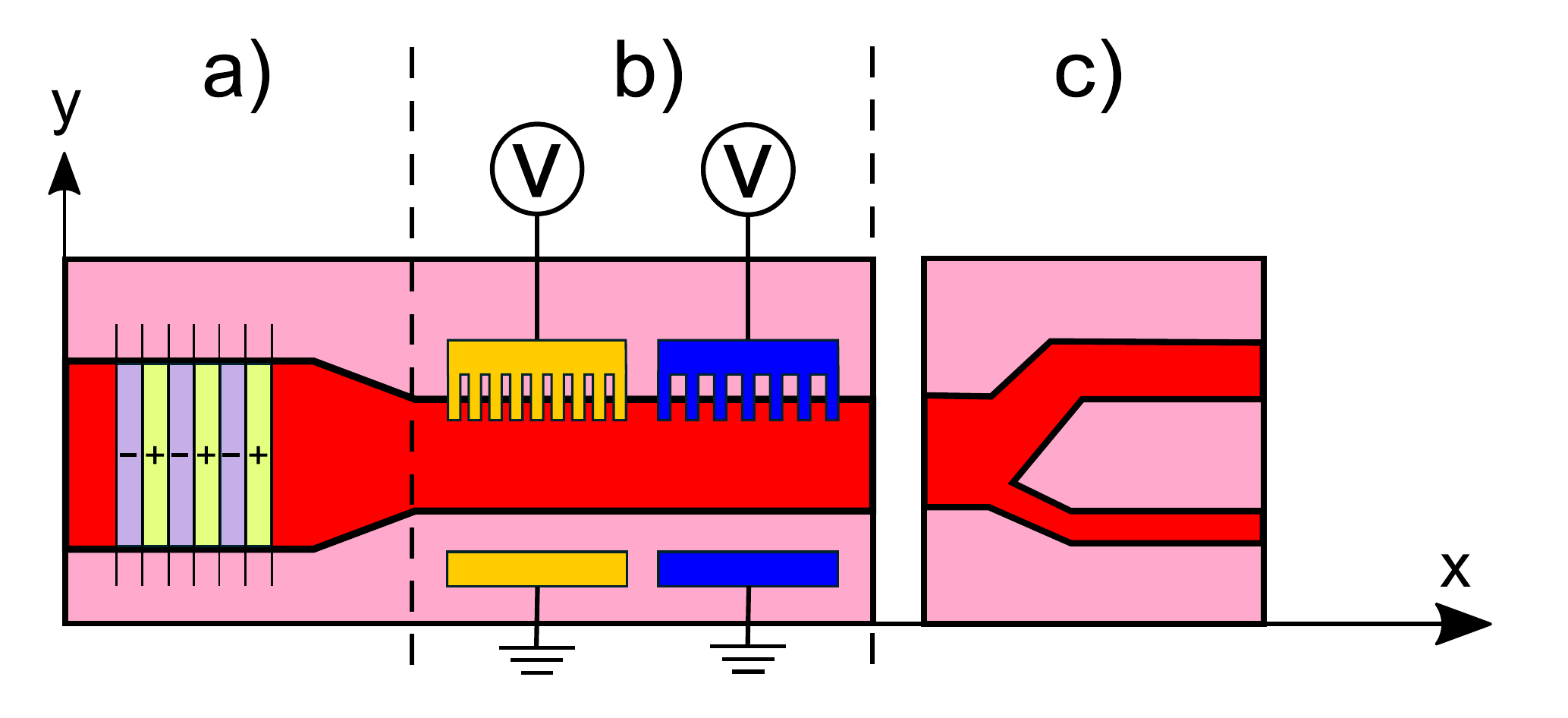}
\caption{ Top view of the waveguide design for generating and testing of spatially mode entangled state. Region (a) is a multimode channel waveguide with a single QPM grating for generating spatially mode entangled states and region (b) is a two-moded channel waveguide with two asymmetric electrode patterns of different periods to independently control coupling among the $H$ and $V$ polarized spatial modes. Part (c) is an asymmetric Y-splitter separating modes $00$ and $10$ into distinct output ports.}
\label{Fig:2}
\end{figure}

In our case the two photons are travelling along the same direction in the waveguide. Therefore, the two mode waveguide needs to be designed in such a way that the two spatial frequencies are sufficiently different so that the coupling among the horizontally polarized spatial modes and vertically polarized spatial modes can be controlled independently. Since the two modes that need to be coupled are of opposite parity, we would need an asymmetric electrode pattern as shown in Fig.~\ref{Fig:2} that would generate an asymmetric electric field distribution which can then enable coupling among the two different parity modes via the electro-optic effect. Because practical waveguides are weakly guiding, the modal electric fields of the $H (V)$-polarized modes (or $y~(z)$-polarized modes) are primarily directed along the $y~(z)$ – direction. Thus the electro-optic coefficient  $r_{23}~(r_{33})$ can be employed to induce coupling among the two spatial modes through the $E_{z}$ component of the applied field.

\begin{figure}[b]
\centering\includegraphics[width=9cm]{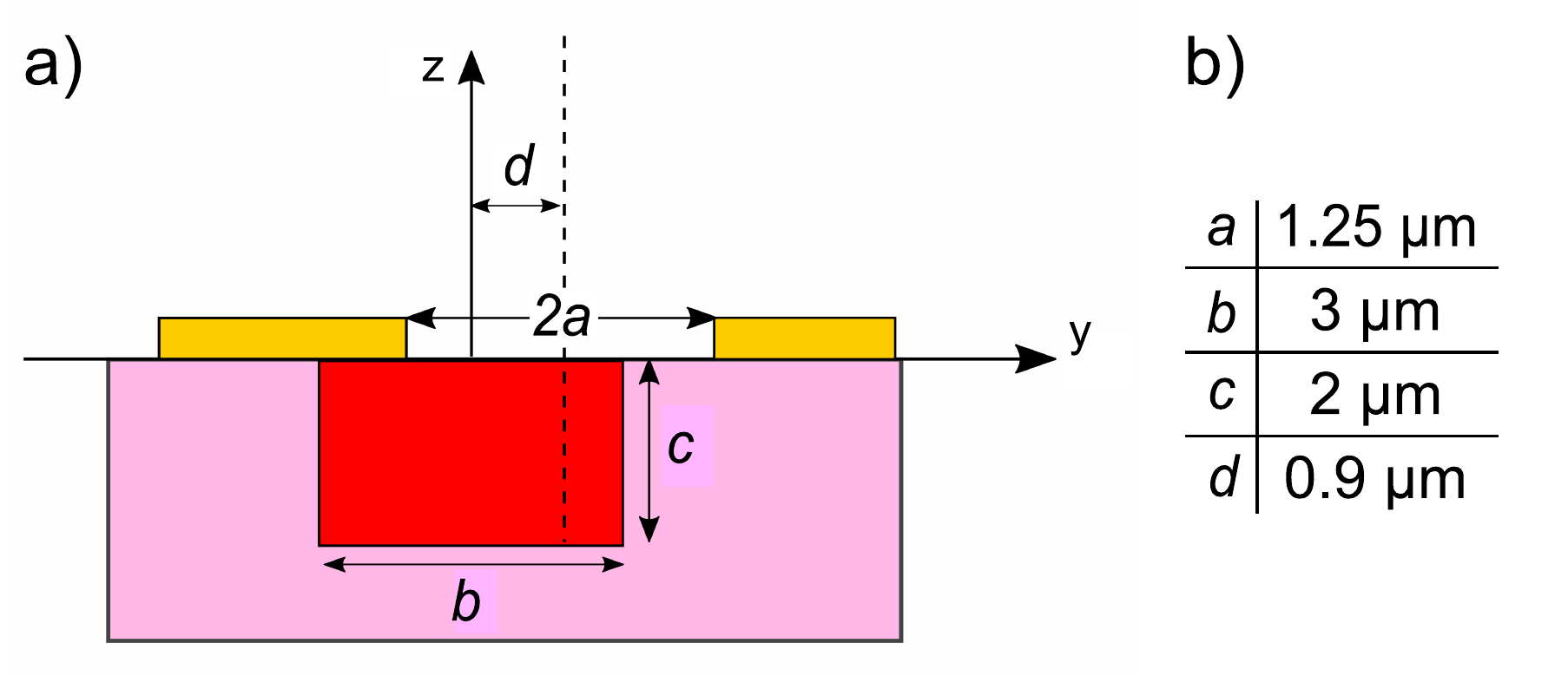}
\caption{(a) Cross-sectional view of the region Fig. 2(b) depicting the geometry of the device. Here, $2a$ is the separation between the two electrodes, $b$ and $c$ corresponds to the width and the depth of the two mode channel waveguide respectively and $d$ is the separation of midpoint of two electrodes (dashed line) with respect to the fixed center of the waveguide. (b) Optimized parameters of the device ensuring an efficient and controlable coupling between the 10 and 00 modes, while preserving both two-mode operation as well as polarization selectivity.}
\label{Fig:3}
\end{figure}

The mixing among the spatial modes can be described through standard coupled mode theory \cite{18,19,20}. Thus in the presence of the periodic electrode pattern of length $L$, the input and output amplitudes of the horizontally polarized spatial modes are related by the following equation:

\begin{equation}
\left(\begin{array}{c}
A_{H}(L)\\
B_{H}(L)
\end{array}\right)=\left(\begin{array}{cc}
\cos(\kappa_{H}L) & \sin(\kappa_{H}L)\\
-\sin(\kappa_{H}L) & \cos(\kappa_{H}L)
\end{array}\right)\left(\begin{array}{c}
A_{H}(0)\\
B_{H}(0)
\end{array}\right),
\end{equation}
where $A$ and $B$ represent respectively the amplitudes of the 00 and 10 mode, while $\kappa_{H}$ is the coupling coefficient between the two spatial modes corresponding to the horizontal polarization. The value of $\kappa_{H}$ can be controlled by changing the applied voltage. Note that this is very similar to the case of polarization splitter where the quantity $\theta = \kappa_{H} L$ would be replaced by the angle made by the principal axes of the polarization splitter with the eigen polarization axes. Similarly the input and output amplitudes of the vertical polarized modes are related by the following equation:

\begin{equation}
\left(\begin{array}{c}
A_{V}(L)\\
B_{V}(L)
\end{array}\right)=\left(\begin{array}{cc}
\cos(\kappa_{V}L) & \sin(\kappa_{V}L)\\
-\sin(\kappa_{V}L) & \cos(\kappa_{V}L)
\end{array}\right)\left(\begin{array}{c}
A_{V}(0)\\
B_{V}(0)
\end{array}\right).
\end{equation}
Exactly similar to the case of horizontal polarization the value of $\kappa_{V}$ can be controlled by an appropriate choice of the applied voltage.

If $\kappa_{H}$ and $\kappa_{V}$ (which are proportional to the applied voltages) represent the coupling coefficients between the 00 and 10 modes corresponding to horizontal and vertical polarizations, then according to coupled mode theory \cite{18} the power coupled among the 00 and 10 modes over a length $L$ are given by $\cos^{2}(\kappa_{H} L)$   and $\cos^{2}(\kappa_{V} L)$   respectively.  A detailed analysis will be discussed in the next section.

The two transverse spatial modes can be separated into two different output paths by using an asymmetric Y-splitter shown in Fig.~\ref{Fig:2}(c) in which both the output waveguides are single-moded but with different propagation constants (which can be easily obtained by choosing different widths of the two waveguides). In such a device, the fundamental symmetric mode at both the signal and the idler wavelengths will exit from the upper waveguide while the first excited antisymmetric mode will exit from the lower waveguide having propagation constant less than the upper waveguide. This way the spatial modes in which the photons are generated can be separated into distinct output ports and individually detected.

Analogously to polarization entanglement, the CHSH inequality is tested by switching randomly between two measurements, characterized by angles $\theta_1$ and $\theta_1'$ for the horizontally polarized photon, and $\theta_2$ and $\theta_2'$ for the vertically polarized photon. Let $P_{mn}(\theta_1, \theta_2)$ denote the coincidence probability of detecting jointly the photon $H$ in the mode $m0$ and the photon $V$ in the mode $n0$, where $m,n=0,1$, for settings $\theta_1 = \kappa_{H} L$ and $\theta_2 = \kappa_{V} L$. These probabilities are used to calculate the correlation function

\begin{equation}
E(\theta_{1},\theta_{2}) = P_{00}(\theta_{1},\theta_{2}) + P_{11}(\theta_{1},\theta_{2}) - P_{01}(\theta_{1},\theta_{2}) - P_{10}(\theta_{1},\theta_{2}).
%\tag{2a}
\label{Eq:Etheta1theta2}
\end{equation}
The correlation functions for all four combinations of detection settings enter the CHSH expression
\begin{equation}
S = E(\theta_{1},\theta_{2}) + E(\theta'_{1},\theta_{2}) + E(\theta'_{1},\theta'_{2}) - E(\theta_{1},\theta'_{2}),
\label{Eq:CHSHcombination}
 %\tag{2b}
\end{equation}
which for separable states is bounded between $-2 \le S \le 2$. For a maximally entangled state it reaches $S = 2\sqrt{2} \approx 2.82843$, which is attained for angles $\theta_{1} = 90^{\circ}, \theta_{2} = 22.5^{\circ} , \theta'_{1} = 45^{\circ}, \theta'_{2} = 67.5^{\circ}$.
As will be shown in Sec.~\ref{Sec:Analysis} these specific values can be obtained by adjusting the voltage applied on the electrodes.

\section{Quantitative analysis}
\label{Sec:Analysis}

In this section we will give detailed analysis for coupling the horizontally (vertically) polarized modes in a two moded channel waveguide and we will show how this leads to testing of CHSH inequality for spatially mode entangled state.

As shown in Figs.~\ref{Fig:2} and \ref{Fig:3} we consider a two moded channel waveguide of width $b$ and depth $c$ with two asymmetric periodic electrode patterns of different periods for coupling the two spatial modes (00 and 10) of horizontal ($H$) polarization and two spatial modes (00 and 10) of vertical ($V$) polarization respectively. We can induce coupling between horizontally (or vertically) polarized modes by periodically modulating the refractive index using the electro-optic effect.

For KTP, in the presence of an applied electric field having components $E_{x}, E_{y}$ and $E_{z}$, the equation of the index ellipsoid will be given by \cite{20,elipsoid}:

\begin{equation}
\frac{x^{2}}{(n_{x}+\Delta\widetilde{n}_{x})^{2}}+\frac{y^{2}}{(n_{y}+\Delta\widetilde{n}_{y})^{2}}
+\frac{z^{2}}{(n_{z}+\Delta\widetilde{n}_{z})^{2}}+2yzr_{42}E_{y}+2xzr_{51}E_{x}=1,
\label{Eq:Ellipsoid}
%%%% \tag{3}
\end{equation}
where $\Delta \widetilde{n}_{x} \approx -n^{3}_x r_{13} E_{z}/2$, $\Delta \widetilde{n}_{y} \approx -n^{3}_y r_{23} E_{z}/2$, $\Delta \widetilde{n}_{z} \approx -n^{3}_z r_{33} E_{z}/2$; $r_{13},r_{23},r_{33},r_{42}$ and $r_{51}$ are  coefficients of electro-optic tensor of KTP which is given as:

\begin{equation}
\left(\begin{array}{ccc}
0 & 0 & r_{13}\\
0 & 0 & r_{23}\\
0 & 0 & r_{33}\\
0 & r_{42} & 0\\
r_{51} & 0 & 0\\
0 & 0 & 0
\end{array}\right). \nonumber
\end{equation}
Assuming a channel waveguide in $z$-cut KTP, with propagation direction along the $x$-direction,  it can be seen from Eq.~(\ref{Eq:Ellipsoid}) that for the $H$-polarized modes (or $y$-polarized modes) whose modal electric field is primarily directed along the $y$-direction, the electro-optic coefficient $r_{23}$ ($\approx$ 16 pm/V) \cite{elopteffect} will induce changes through the $E_{z}$ component of the applied field. Similarly for the $V$-polarized modes (or $z$-polarized modes) whose modal electric field is primarily directed along the $z$-direction, the electro-optic coefficient $r_{33}$ ($\approx$ 36 pm/V) will induce changes through the $E_{z}$ component of the applied electric field. The term containing $r_{42}$ will try to induce coupling among the $y$- and $z$-polarizations; however since the propagation constants of the two polarized modes are different, it would need a periodic perturbation of a specific period to induce coupling. This period being different from those required for coupling among the two spatial modes, this term is not expected to have any effect. Similarly the effect of the $r_{51}$ term can be neglected.

In the following we analyze how this refractive index modulation will lead to coupling among the two modes in the waveguide. The total electric field in the $y$ ($H$-polarized) or the $z$ ($V$-polarized) direction can be written as a superposition of the two corresponding modes with appropriate $x$-dependent amplitudes:

\begin{flalign}
\Phi_{y(z)}(x,y,z) &=E_{00}^{H(V)}(x,y,z)+E_{10}^{H(V)}(x,y,z)= \nonumber \\
&= A_{H(V)}(x)\psi_{00}^{H(V)}(y,z)e^{i(\omega t-\beta_{00}^{H(V)}x)}+B_{H(V)}(x)\psi_{10}^{H(V)}(y,z)e^{i(\omega t-\beta_{10}^{H(V)}x)}, %%\tag{4}
\end{flalign}
where $E_{00}^{H(V)}$ and $E_{10}^{H(V)}$ is the electric field distribution of $H(V)$-polarized 00 and 10 modes respectively. The two mode field profiles satisfy the following equations:
\begin{subequations}
\begin{flalign}
\frac{\partial^{2}\psi_{00}^{H(V)}}{\partial y^{2}}+\frac{\partial^{2}\psi_{00}^{H(V)}}{\partial z^{2}}+[k_{0}^{2}n_{y(z)}^{2}(y,z)-(\beta_{00}^{H(V)})^{2}]\psi_{00}^{H(V)}=0
%%% \tag{5a}
\\
 \frac{\partial^{2}\psi_{10}^{H(V)}}{\partial y^{2}}+\frac{\partial^{2}\psi_{10}^{H(V)}}{\partial z^{2}}+[k_{0}^{2}n_{y(z)}^{2}(y,z)-(\beta_{10}^{H(V)})^{2}]\psi_{10}^{H(V)}=0.
%%% \tag{5b}
\end{flalign}
\end{subequations}
Here $\psi_{mn}^{H(V)}$ is the field profile of the $H(V)$-polarized $mn$ mode. We assume that the applied electric field creates a refractive index modulation given by:
\begin{equation}
\widetilde{n}_{y(z)}^{2}(x,y,z)\approx n_{y(z)}^{2}(y,z)+2n_{y(z)}(y,z)\Delta n_{y(z)}(x,y,z). \nonumber
\end{equation}
With periodic electrodes of spatial period $\Lambda_{H(V)}$, the generated refractive index modulation becomes
\begin{equation}
\Delta n_{y(z)}(x,y,z)=\Delta\widetilde{n}_{y(z)}(y,z)\sin(K_{H(V)}x), \nonumber
\end{equation}
where $K_{H(V)}=\frac{2\pi}{\Lambda_{H(V)}}$ is the spatial frequency of the grating used for coupling among the $H$-modes~or the $V$-modes. Using standard coupled mode theory \cite{18,19,20} we obtain the following two coupled equations describing the coupling among the two modes:
\begin{subequations}
\begin{flalign}
\frac{\partial A_{H(V)}}{\partial x}=\kappa_{12}^{H(V)}B_{H(V)}e^{i(\Delta\beta_{H(V)}-K_{H(V)})x}
\label{Eq:dAdx}
%%\tag{6a}
\\
\frac{\partial B_{H(V)}}{\partial x}=-\kappa_{21}^{H(V)}A_{H(V)}e^{-i(\Delta\beta_{H(V)}-K_{H(V)})x},
\label{Eq:dBdx}
%%\tag{6b}
\end{flalign}
\end{subequations}
where
\begin{subequations}
\begin{flalign}
\kappa_{12}^{H(V)}=\frac{k_{0}^{2}}{2\beta_{00}^{H(V)}}\int_{-\infty}^{\infty}\int_{-\infty}^{\infty}(\psi_{00}^{H(V)})^{*}n_{y(z)}(y,z)\Delta\widetilde{n}_{y(z)}(y,z)(\psi_{10}^{H(V)})dydz, %%\tag{7a}
\\
\kappa_{21}^{H(V)}=\frac{k_{0}^{2}}{2\beta_{10}^{H(V)}}\int_{-\infty}^{\infty}\int_{-\infty}^{\infty}(\psi_{10}^{H(V)})^{*}n_{y(z)}(y,z)\Delta\widetilde{n}_{y(z)}(y,z)(\psi_{00}^{H(V)})dydz, %%\tag{7b} \\
\end{flalign}
\end{subequations}
and
\begin{equation}
\Delta\beta_{H(V)}=\beta_{00}^{H(V)}-\beta_{10}^{H(V)}.
%\tag{7c}
\end{equation}
Solving Eqs.~(\ref{Eq:dAdx}) and (\ref{Eq:dBdx}), we get the general solution as:
\begin{multline}
\begin{pmatrix}
A_{H(V)}(x)\\
B_{H(V)}(x)
\end{pmatrix} = %%%\tag{8}
\\
\begin{pmatrix}
\left[\cos(\gamma_{H(V)}x)-\frac{i\delta_{H(V)}}{\gamma_{H(V)}}\sin(\gamma_{H(V)}x)\right]e^{i\delta_{H(V)}x},\left[\frac{\kappa_{H(V)}}{\gamma_{H(V)}}\sin(\gamma_{H(V)}x)\right]e^{i\delta_{H(V)}x}\\
\left[-\frac{\kappa_{H(V)}}{\gamma_{H(V)}}\sin(\gamma_{H(V)}x)\right]e^{-i\delta_{H(V)}x},\left[\cos(\gamma_{H(V)}x)+\frac{i\delta_{H(V)}}{\gamma_{H(V)}}\sin(\gamma_{H(V)}x)\right]e^{-i\delta_{H(V)}x}
\end{pmatrix}\begin{pmatrix}
A_{H(V)}(0)\\
B_{H(V)}(0)
\end{pmatrix},
\label{Eq:SolutionAB}
\end{multline}
where:
\begin{flalign*}
\delta_{H(V)}=\frac{\Delta\beta_{H(V)}-K_{H(V)}}{2},~~\kappa_{H(V)}=\sqrt{\kappa_{12}^{H(V)}\kappa_{21}^{H(V)}},~~(\gamma_{H(V)})^{2}=(\kappa_{H(V)})^{2}+(\delta_{H(V)})^{2}.
\end{flalign*}
For the input state given by $\ket{00_{H},10_{V}}$ with the $H$-polarized photon in the 00 mode and the $V$-polarized photon in the 10 mode, the initial conditions at $x = 0$ correspond to:
\begin{subequations}
\begin{equation}
A_{H}(x=0) = 1~~\&~~B_{H} (x=0) = 0;~~B_{V} (x=0) = 1~~\&~~A_{V}(x=0) = 0. %%% \tag{9a}
\label{Eq:InitialAHBV}
\end{equation}
Similarly for the input state given by $\ket{10_{H},00_{V}}$ the initial conditions at $x = 0$ correspond to:
\begin{equation}
B_{H}(x=0) = 1~~\&~~A_{H} (x=0) = 0;~~A_{V} (x=0) = 1~~\&~~B_{V}(x=0) = 0. %%%%\tag{9b}
\label{Eq:InitialAVBH}
\end{equation}
\end{subequations}
Now, owing to the initial condition (\ref{Eq:InitialAHBV}), Eq.~(\ref{Eq:SolutionAB}) becomes:
\begin{subequations}
\begin{flalign}
A_{H}(x)=\left[\cos(\gamma_{H}x)-\frac{i\delta_{H}}{\gamma_{H}}\sin(\gamma_{H}x)\right]e^{i\delta_{H}x};~~B_{H}(x)=\left[-\frac{\kappa_{H}}{\gamma_{H}}\sin(\gamma_{H}x)\right]e^{-i\delta_{H}x} %%%\tag{10a}
\\
A_{V}(x)=\left[\frac{\kappa_{V}}{\gamma_{V}}\sin(\gamma_{V}x)\right]e^{i\delta_{V}x};~~B_{V}(x)=\left[\cos(\gamma_{V}x)+\frac{i\delta_{V}}{\gamma_{V}}\sin(\gamma_{V}x)\right]e^{-i\delta_{V}x}. %%%\tag{10b}
\end{flalign}
\end{subequations}
Thus the powers in modes 00 and 10 for both polarizations ($H$ and $V$) at the end of the interaction of length $L$ are given by:
\begin{flalign*}
P_{H}^{00}(L)=|A_{H}|^{2}=\cos^{2}(\gamma_{H}L)+\left(\frac{\delta_{H}}{\gamma_{H}}\right)^{2}\sin^{2}(\gamma_{H}L);~~P_{H}^{10}(L)=|B_{H}|^{2}=\left(\frac{\kappa_{H}}{\gamma_{H}}\right)^{2}\sin^{2}(\gamma_{H}L) \\
P_{V}^{00}(L)=|A_{V}|^{2}=\left(\frac{\kappa_{V}}{\gamma_{V}}\right)^{2}\sin^{2}(\gamma_{V}L);~~P_{V}^{10}(L)=|B_{V}|^{2}=\cos^{2}(\gamma_{V}L)+\left(\frac{\delta_{V}}{\gamma_{V}}\right)^{2}\sin^{2}(\gamma_{V}L).
\end{flalign*}
Similarly, solving Eq. (\ref{Eq:SolutionAB}) for the initial condition (\ref{Eq:InitialAVBH}), we obtain powers in modes 00 and 10 for both polarizations ($H$ and $V$) at the end of the interaction of length $L$ as:
\begin{flalign*}
P_{H}^{00}(L)=|A_{H}|^{2}=\left(\frac{\kappa_{H}}{\gamma_{H}}\right)^{2}\sin^{2}(\gamma_{H}L);~~P_{H}^{10}(L)=|B_{H}|^{2}=\cos^{2}(\gamma_{H}L)+\left(\frac{\delta_{H}}{\gamma_{H}}\right)^{2}\sin^{2}(\gamma_{H}L) \\
P_{V}^{00}(L)=|A_{V}|^{2}=\cos^{2}(\gamma_{V}L)+\left(\frac{\delta_{V}}{\gamma_{V}}\right)^{2}\sin^{2}(\gamma_{V}L);~~P_{V}^{10}(L)=|B_{V}|^{2}=\left(\frac{\kappa_{V}}{\gamma_{V}}\right)^{2}\sin^{2}(\gamma_{V}L).
\end{flalign*}
We show in Sec.~4 that it is possible to choose the grating period corresponding to the signal and idler wavelengths to satisfy the conditions $\delta_{H(V)} = 0$; in such a case we have:
\begin{equation}
\left(\begin{array}{c}
A_{H(V)}(L)\\
B_{H(V)}(L)
\end{array}\right)=\left(\begin{array}{cc}
\cos(\kappa_{H(V)}L) & \sin(\kappa_{H(V)}L)\\
-\sin(\kappa_{H(V)}L) & \cos(\kappa_{H(V)}L)
\end{array}\right)\left(\begin{array}{c}
A_{H(V)}(0)\\
B_{H(V)}(0)
\end{array}\right).
%%%%\tag{11}
\end{equation}
Thus, separating and detecting photons in the modes $00$ and $10$ after the rotation corresponds to realizing a projection in the spatial modal basis given by:
\begin{subequations}
\begin{align}
\ket{{\theta}_{H(V)}} & = \cos\theta \ket{00_{H(V)}} + \sin\theta \ket{10_{H(V)}} %%%%\tag{12a}
\\
\ket{({\theta + {\textstyle\frac{\pi}{2}}})_{H(V)}} & = - \sin\theta \ket{00_{H(V)}} + \cos\theta \ket{10_{H(V)}}.
%%% \tag{12b}
\end{align}
\end{subequations}
For a general two-photon state described by a density matrix $\hat{\varrho}$, the coincidence probabilities are given by expressions
\begin{equation}
P_{mn}(\theta_1, \theta_2) = \bra{(\theta_1+ m {\textstyle\frac{\pi}{2}})_H , (\theta_2 + n {\textstyle\frac{\pi}{2}})_V}
\hat{\varrho}
\ket{(\theta_1 + m {\textstyle\frac{\pi}{2}})_H , (\theta_2+ n {\textstyle\frac{\pi}{2}})_V},
\end{equation}
where $m,n=0,1$.
Using these expressions we can evaluate the correlation functions according to Eq.~(\ref{Eq:Etheta1theta2}) and further the CHSH combination $S$ defined in Eq.~(\ref{Eq:CHSHcombination}).

\section{Numerical results}
\label{Sec:Numerical}

For generation of spatially mode entangled pairs of photons we consider an ion-diffused potassium titanyl phosphate (KTP) channel waveguide with a step-index profile. The values of the KTP substrate refractive index ($n_{s}$) for different wavelengths and different polarizations were calculated using Sellmeier equations given in \cite{21}. The refractive index difference ($\Delta n$) for a waveguide is taken to be 0.02.
\begin{figure}[hbt]
\centering\includegraphics[width=12cm]{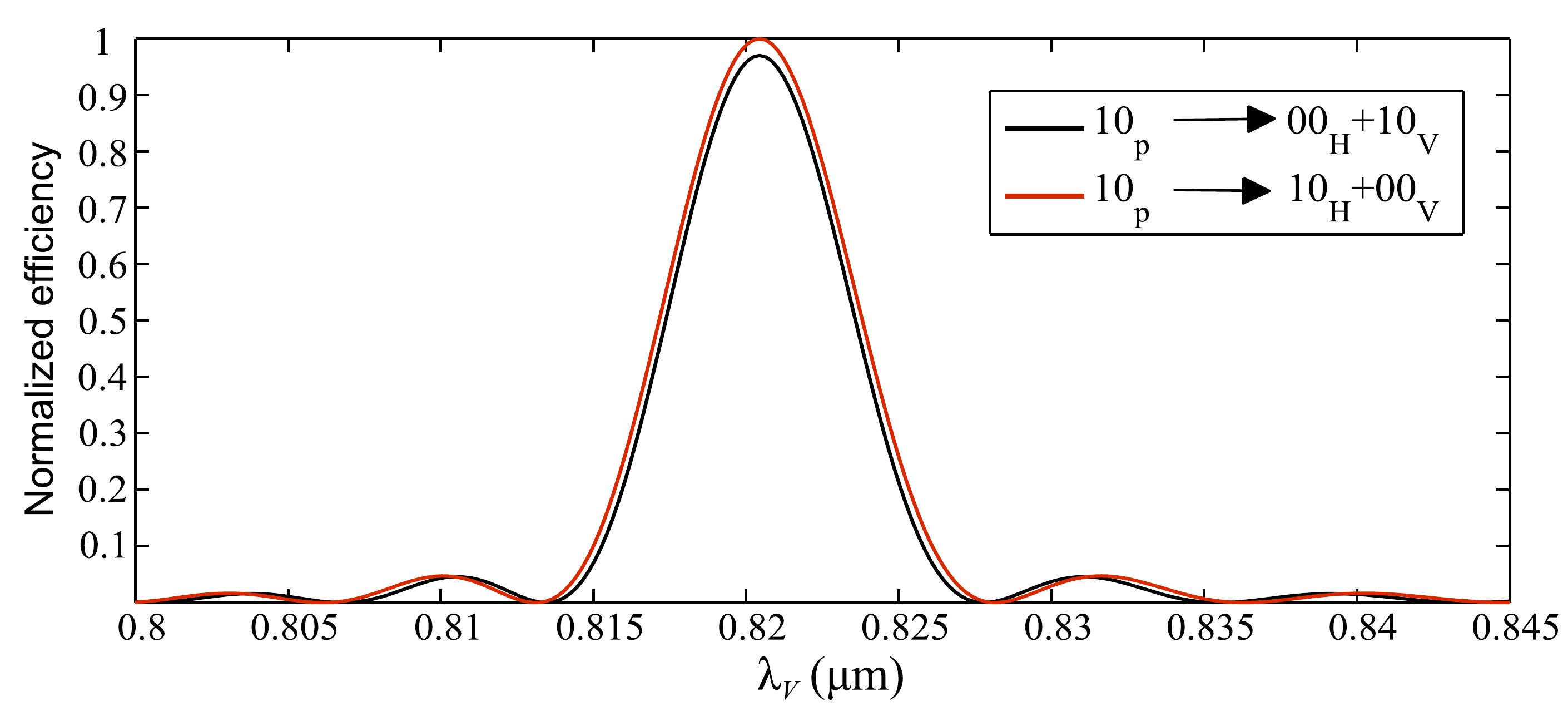}
\caption{Spectral distributions of the two SPDC processes used to prepare a pair of photons entangled
in transverse spatial modes.}
\label{Fig:4}
\end{figure}
Carrying out the simulations we obtain the optimized waveguide parameters as waveguide width = 5.0 $\mathrm{\mu m}$, waveguide depth = 2.0 $\mathrm{\mu m}$, QPM grating period = 6.956 $\mathrm{\mu m}$, length of waveguide = 1 mm; for these parameters the QPM conditions for both the processes can be simultaneously satisfied. Figure~\ref{Fig:4} depicts the spectral dependence of the two SPDC processes, including conversion efficiencies given by spatial overlaps of triplets of transverse modes involved in each process according to Eq.~(\ref{Eq:TwoProcesses}).
It is seen that spatially mode entangled pairs of photons can be generated with the maximum probability attained for the wavelengths ($\lambda_{p},\lambda_{H},\lambda_{V}$) = (392~nm, 750.776~nm, 820.435~nm). Here, $\lambda_{p},\lambda_{H}$ and $\lambda_{V}$ corresponds to pump, signal and idler wavelength respectively. Taking into account slightly unequal conversion efficiencies of the two processes and residual spectral distinguishability of the generated pairs, the generated two-photon state can be modelled by a normalized density matrix
\cite{ProgOptReview}:
\begin{multline}
\hat{\varrho} = w \proj{00_H , 10_V} + (1-w) \proj {10_H , 00_V} \\
+ v \bigl( \ket{00_H , 10_V} \bra{10_H , 00_V} + \ket{10_H , 00_V} \bra{00_H , 10_V} \bigr)
\label{Eq:RealisticRho}
\end{multline}
where $w=0.4825$ and $v=0.4979$. Here we have neglected photons in other spatial mode pairs that could be generated by SPDC in a multimode waveguide \cite{KarpinskiSPIE, JachuraOpEx2014}. They can be removed either by spatial filtering during the passage to the two-mode waveguide depicted in
Fig.~\ref{Fig:2}(b) or by coarse spectral filtering thanks to the spatial-spectral correlations present in SPDC in multimode waveguides \cite{MosleyChristPRL09,KarpinskiSPIE, JachuraOpEx2014}.

\begin{figure}[b!]
\centering\includegraphics[width=11cm]{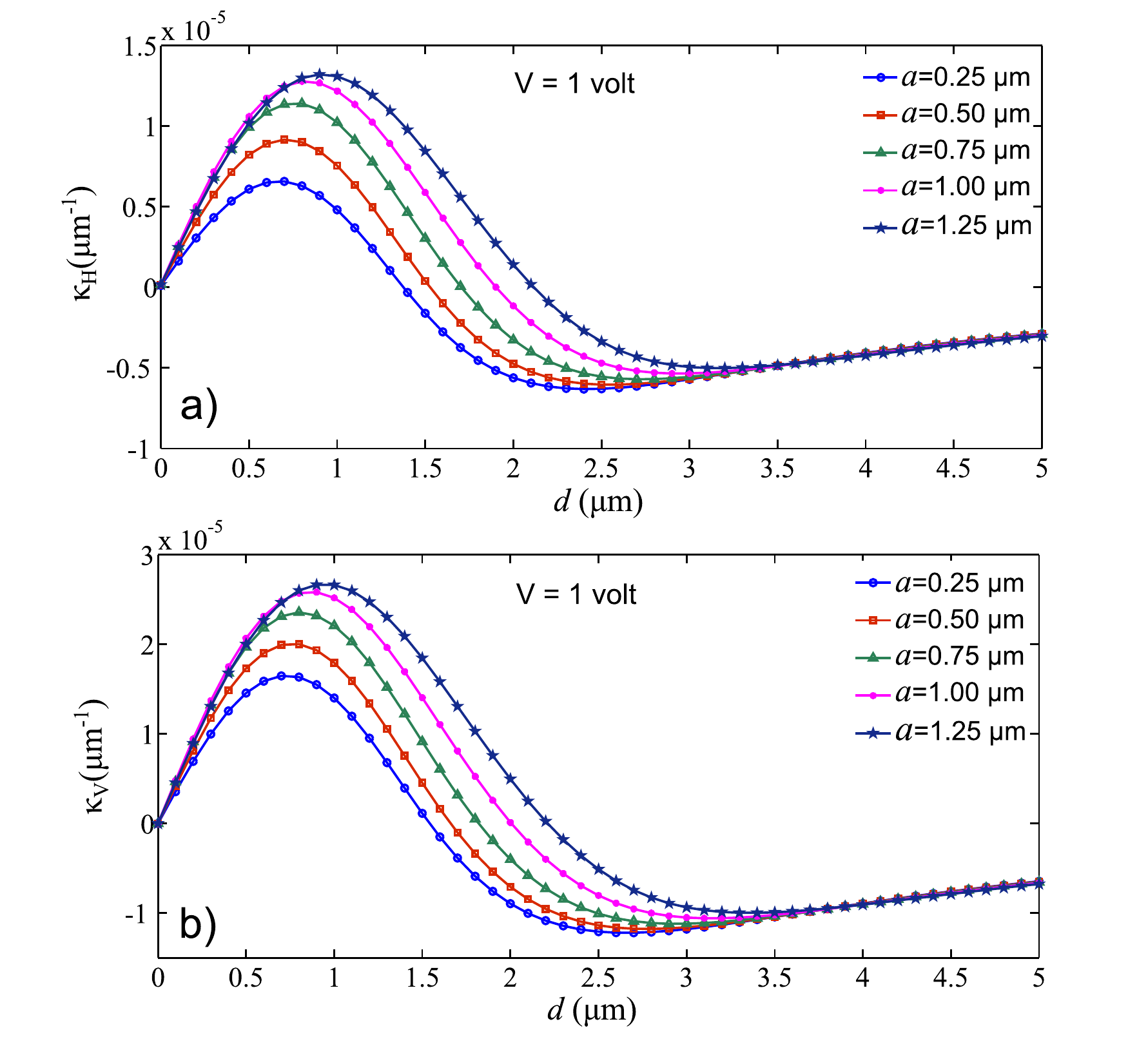}
\caption{(a) Variation of $\kappa_{H}$ with $d$ for 1 volt for different value of $a$, (b) Variation of $\kappa_{V}$ with $d$ for 1 volt for different value of $a$.}
\label{Fig:6}
\end{figure}

In order to perform rotations in the spatial mode bases of the photons $H$ and $V$, the width of the waveguide is reduced from 5 $\mathrm{\mu m}$ to 3 $\mathrm{\mu m}$ (see Fig. 2) so, that it supports only two spatial modes at signal and idler wavelengths (i.e. 00 and 10 mode). The purpose of reducing the waveguide width is twofold: first it will act like a modal filter to remove modes other than the 00 and 10 modes and secondly it will introduce significant difference between spatial frequencies $K_{H}=(\beta_{00}^{H}-\beta_{10}^{H})$ and $K_{V}=(\beta_{00}^{V}-\beta_{10}^{V})$ so that the grating used for introducing coupling between the horizontally polarized modes does not introduce coupling between vertically polarized modes and vice versa.

Fig.~\ref{Fig:3} shows the cross section of the channel waveguide and the electrode patterns for coupling among the $H$-polarized modes and the $V$-polarized modes.  The transverse dependence of the $z$-component of the electric field distribution is given by \cite{22}:
\begin{equation}
E_{z}(y,z)=\frac{V}{\pi\sqrt{2}}\frac{[-(a^{2}+z^{2}-y^{2})+\{a^{4}+y^{4}+z^{4}+2y^{2}z^{2}+2a^{2}(z^{2}-y^{2})\}^{1/2}]^{1/2}}{[a^{4}+y^{4}+z^{4}+2y^{2}z^{2}+2a^{2}(z^{2}-y^{2})]^{1/2}}, \tag{21}
\end{equation}
where $2a$ is a distance between two electrodes and $V$ is the voltage applied on the electrodes.

Numerical optimization of the electrode positions vis a vis the waveguide leads to the maximal mode coupling strength for the electrodes positioned asymmetrically as shown in Fig.~\ref{Fig:3}. Figure~\ref{Fig:6} shows the variation of the coupling coefficients $\kappa_{H},\kappa_{V}$ for different values of half-separation between electrodes  $a$ as a function of the distance between electrodes midpoint and fixed waveguide center $d$ (see Fig.~3(a)) for an applied voltage of $1~\mathrm{V}$. It can be seen from Fig.~\ref{Fig:6}(a) and Fig.~\ref{Fig:6}(b) that for the optimum position of electrodes (see Fig.~3(b)) value of $\kappa_{H}$ and $\kappa_{V}$ found to be $1.32\times 10^{-5}~\mathrm{\mu m^{-1}}$ and $2.66\times10^{-5} ~\mathrm{\mu m^{-1}}$ respectively for an applied voltage of $1~\mathrm{V}$.

The periods of the electrodes are chosen such that the spatial frequency $K$ of the index modulation matches with $\Delta\beta$ of the corresponding pair of modes that are to be coupled. For the waveguides used in the simulation the respective grating periods are $\Lambda_{H} = 102.5~\mathrm{\mu m} $ and $\Lambda_{V} = 105.1~\mathrm{\mu m} $ which are significantly different from each other.

\begin{figure}[h!]
\centering\includegraphics[width=12cm]{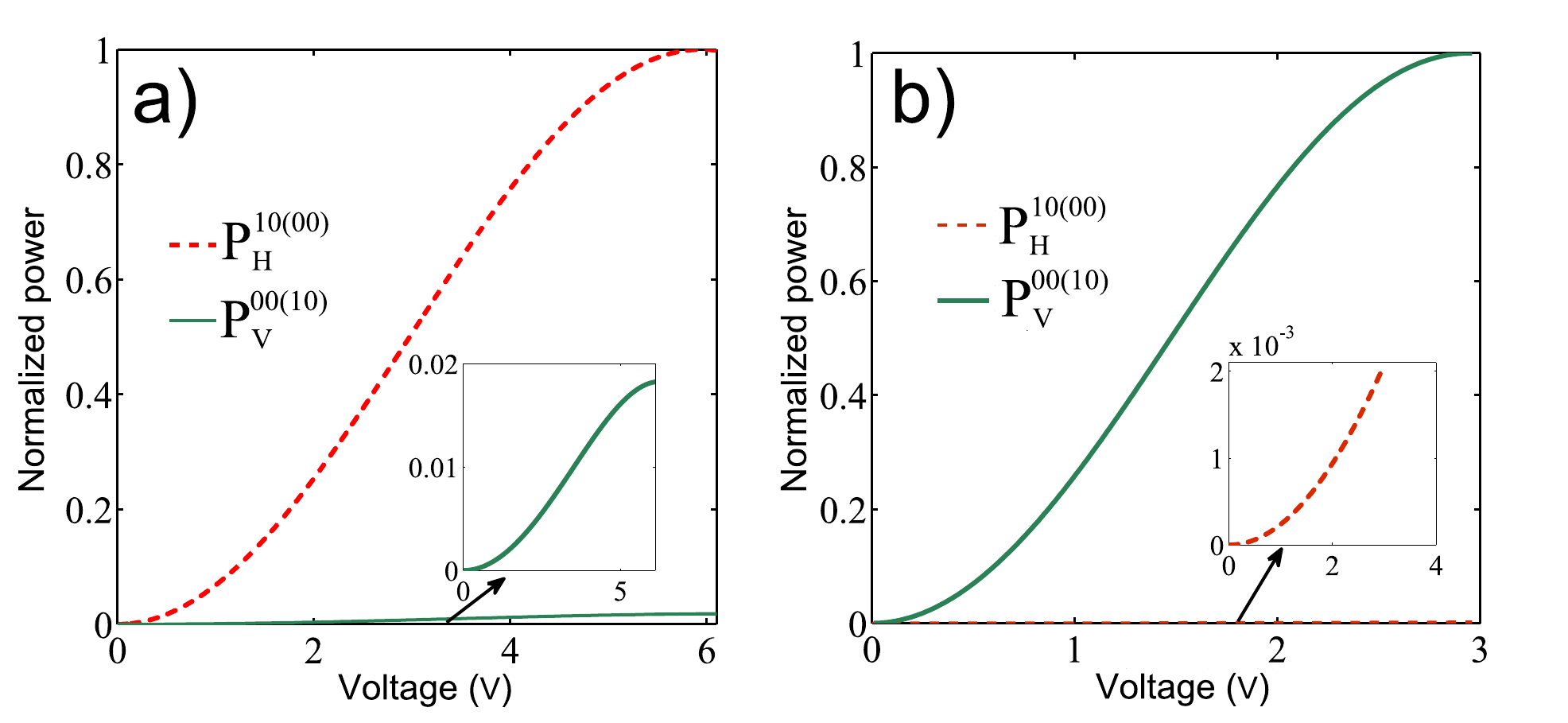}
\caption{Variation of normalized power for vertically polarized 00(10) and horizontally polarized 10(00) mode with voltage, corresponding to the spatial frequency (a) $K_{H} = 2\pi/102.5 \mu m$ and (b) $K_{V} = 2\pi/105.1 \mu m$.}
\label{Fig:7}
\end{figure}

In order to check whether the grating that induces coupling among the $H$-polarized modes also induces coupling among the $V$-polarized modes, in Fig.~\ref{Fig:7}(a) we have plotted the variation of the coupled power of  $H$ and $V$ polarizations for an electrode length of 2 cm. As can be seen in this case although the entire power in the $H$-polarization can be coupled into the other spatial mode, less than 2\% of the light gets coupled in the other polarization. This is primarily due to non-satisfaction of quasi phase matching condition (see Fig.~\ref{Fig:7}(a)). Similarly spatial frequency $K_{V}$ can induce coupling among the $V$-polarized 00 and 10 modes without inducing much coupling among the 00 and 10 modes of $H$-polarization as shown in Fig.~\ref{Fig:7}(b).

Because of small deviations of the generated two-photon state given in Eq.~(\ref{Eq:RealisticRho}) from the maximally entangled form, we performed a numerical search of measurement settings that yields the strongest violation of the CHSH inequality. The search produced values $\theta_1 =87.850^{\circ}$, $\theta_2= 24.598^{\circ}$, $\theta_1'=42.832^{\circ}$, and $\theta_2'= 69.720^{\circ}$ with the resulting value of the CHSH combination equal to $S=2.82236$.

\section{Conclusions}
\label{Sec:Conclusions}

In integrated photonic platforms, spatial multiplexing is being actively explored as means to increase the information capacity of optical links. Such an approach can also prove fruitful in implementations of quantum-enhanced technologies exploiting superpositions of multiple spatial modes. In this paper we presented a design for integrated electro-optic devices that enables deterministic transformations of mode superpositions in a two-dimensional subspace. We discussed application of such devices to verify entanglement generated in transverse spatial modes by testing the CHSH inequality. Fabrication of components described here should be within the reach of current manufacturing capabilities \cite{23}.

\section*{Acknowledgments}
This research was partly supported by the EU 7th Framework Programme projects SIQS (Grant Agreement No. 600645) and PhoQuS@UW (Grant Agreement No. 316244).

\end{document}